\documentstyle[11pt,equation,
epsfig]{article}
\begin{document}
\newcommand{\Lt}{\tilde{\Lambda}}
\newcommand{\Ct}{\tilde{C}}
\newcommand{\phit}{\tilde{\phi}}
\newcommand{\done}{\delta\phi_1}
\newcommand{\dy}{\delta y}
\newcommand{\tr}{\rm Tr}
\newcommand{\sx}{\sigma}
\newcommand{\mpl}{m_{Pl}}
\newcommand{\Mpl}{M_{Pl}}
\newcommand{\lx}{\lambda}
\newcommand{\Lx}{\Lambda}
\newcommand{\kx}{\kappa}
\newcommand{\ex}{\epsilon}
\newcommand{\be}{\begin{equation}}
\newcommand{\ee}{\end{equation}}
\newcommand{\een}{\end{subequations}}
\newcommand{\ben}{\begin{subequations}}
\newcommand{\beq}{\begin{eqalignno}}
\newcommand{\eeq}{\end{eqalignno}}
\def \lta {\mathrel{\vcenter
     {\hbox{$<$}\nointerlineskip\hbox{$\sim$}}}}
\def \gta {\mathrel{\vcenter
     {\hbox{$>$}\nointerlineskip\hbox{$\sim$}}}}
\pagestyle{empty}
\noindent
\begin{flushright}
December 2000
\\
\end{flushright} 
\vspace{3cm}
\begin{center}
{ \Large \bf
On Brane Stabilization and the Cosmological Constant
} 
\\ \vspace{1cm}
{\Large 
N. Tetradis 
} 
\\
\vspace{1cm}
{\it
Department of Physics, University of Crete,
710 03 Heraklion, Crete, Greece
} 
\\
and
\\
{\it
Department of Physics, University of Athens,
157 71 Athens, Greece
} 
\\
\vspace{3cm}
\abstract{
We address the problem of the cosmological constant within the
Randall-Sundrum scenario with a brane stabilization mechanism.
We consider brane tensions of general form. We examine the
conditions under which a small change of the positive tension of
the first brane can be absorbed in a small modification of the
two-brane configuration, instead of manifesting itself as a
cosmological constant. We demonstrate that it is possible to have
a cosmological constant in the range predicted by recent observational
data, if there is an ultraviolet 
cutoff of order 10 TeV in the contributions to
the brane tension from vacuum fluctuations.
\\
\vspace{1cm}
} 
\end{center}
\vspace{4cm}
\noindent

\newpage

\pagestyle{plain}
\setcounter{page}{1}

\setcounter{equation}{0}

\paragraph{Introduction:}
The smallness of the cosmological constant has defied explanation
despite a long history of attempts \cite{weinberg}. The problem is
further complicated by the recent observational evidence that the present 
cosmological constant may
be of the order of the critical density of the Universe \cite{obsc}.
An interesting proposal for the resolution of the problem was made in ref.
\cite{rubshap}: If our four-dimensional world is embedded in 
a higher-dimensional space-time,
the effect of non-zero vacuum energy 
may affect only the curvature in the extra dimensions, allowing for 
a flat four-dimensional metric. 

The possible existence of large
extra dimensions \cite{larged,akama,savas} provides a setup in which
to realize the above idea. The
Standard Model fields are assumed to be localized on a four-dimensional
surface, 
the ``brane'', and only gravitons can propagate in the ``bulk'' of the
extra dimensions \cite{savas}. For a
class of geometries characterized as ``warped'' the low energy gravitons
are also localized on the brane \cite{rs1}. This framework provides a new
opportunity to confront the cosmological constant problem and several 
attempts have been made in this direction \cite{adks}-\cite{other}.
We are interested in the possibility that changes of the vacuum
energy of the brane, the brane tension, can be absorbed in modifications
of the bulk geometry so that the effective four-dimensional constant
remains zero or almost zero. Existing scenarios have undesirable features,
such as the presence of naked singularities in the metric, or very specific
assumptions about the form of the couplings in the effective action of
the theory \cite{adks}.

Our setup is that of ref. \cite{rs1}, as generalized in refs. 
\cite{goldwise,dewolfe}: We consider a five-dimensional
system with two four-dimensional branes, of positive
and negative tension respectively. These are located at the boundaries of  
a compact fifth dimension with anti-deSitter bulk metric. 
The geometry is warped, in a way that the
low-energy gravitons are localized near the positive-tension brane. 
The distance between the two branes is not arbitrary as in ref. \cite{rs1}.
The presence of a bulk field with a non-trivial potential permits only static
configurations with specific values for the size of the fifth dimension
\cite{goldwise}. The backreaction of the field on the metric is taken into
account along the lines of ref. \cite{dewolfe}.

We consider brane tensions of general form, which also depend on
the value of the bulk field at the location of the brane.
Our starting point is a configuration with zero four-dimensional
cosmological constant.
We then discuss under what conditions small modifications of the
positive tension of the first brane, which we identify with our
low-energy world, can be absorbed into small displacements of the
branes, instead of manifesting themselves as a cosmological constant.

\paragraph{The fine-tuning:}
We consider a system of two branes in the background of
a bulk scalar field $\phi$. The action is given by 
\be
S=\int d^4x\,dy \, \sqrt{\left| \det g_{\mu\nu}\right|} \,
\left[ - 2 M^3 R + \frac{1}{2} \left( \partial \phi \right)^2
- V(\phi) \right] -\sum_{\alpha=1,2} 
\int d^4x \, \sqrt{\left| \det g_{ij}\right|}\, \lx_\alpha(\phi).
\label{action0}
\ee
By rescaling all dimensionful
quantities by $2M$ (which is of the order of the fundamental Planck's constant)
we obtain
\be
S=\int d^4x\,dy \, \sqrt{\left| \det g_{\mu\nu}\right|} \,
\left[ -\frac{1}{4} R + \frac{1}{2} \left( \partial \phi \right)^2
- V(\phi) \right] -\sum_{\alpha=1,2} 
\int d^4x \, \sqrt{\left| \det g_{ij}\right|}\, \lx_\alpha(\phi),
\label{action}
\ee
consistently with the notation of ref. \cite{dewolfe}. 
We emphasize that all quantities in eq. (\ref{action}) are dimensionless,
even though they are denoted by the same symbols as in eq. (\ref{action0}).

For the metric we assume the ansatz 
\be
ds^2 = e^{2 A(y)}\, \eta_{ij} \,dx^i dx^j - dy^2
\label{metr1} \ee
with space-time topology $R^{3,1}\times S^1/Z_2$ \cite{rs1}. 
The two branes are located at the boundaries of the fifth dimension.
Einstein's equations and the equation of motion of the field are
\cite{dewolfe}
\beq 
\phi''+4 A' \phi'= &\frac{\partial V(\phi)}{\partial \phi} +
\sum_{\alpha=1,2} 
\frac{\partial \lx_\alpha (\phi)}{\partial \phi}\, \delta (y-y_\alpha)
\label{eoma1} \\
A'' = &-\frac{2}{3}\phi'^2-\frac{2}{3}
\sum_{\alpha=1,2} 
\lx_\alpha (\phi) \,\delta (y-y_\alpha)
\label{eoma2} \\
A'^2 = &-\frac{1}{3} V(\phi) + \frac{1}{6} \phi'^2.
\label{eoma3}
\eeq
Primes denote derivatives with respect to $y$.
We set $y_1=0$ and $y_2=R$.

The solutions of the above equations for general potentials $V(\phi)$
predict a fixed distance $R$ between the two branes. They generalize
the stabilization mechanism of ref. \cite{goldwise} by taking into account
the backreaction of the scalar field on the gravitational background.
The functions $\lx_\alpha (\phi)$ are characterized as the brane tensions. 
Their form is determined by the vacuum energy of the fields that live
on the brane. We have also allowed for 
an interaction of these fields with the bulk field, so
that the tensions depend on $\phi$.
The presence of the branes
imposes boundary conditions for $A'(y)$ and $\phi(y)$ at $y=0,R$.
The integration of eq. (\ref{eoma1}), (\ref{eoma2})
around the $\delta$-functions and use of the $Z_2$ symmetry 
leads to
\beq 
y=0
~~~~~~~~~~~~~~~~
\phi'=&~\frac{1}{2}\frac{\partial \lx_1(\phi)}{ \partial \phi} 
~~~~~~~~~~~~~~~~
A'=-\frac{1}{3} \lx_1(\phi)
\label{bound1} \\
y=R
~~~~~~~~~~~~~~~~
\phi'=&~-\frac{1}{2}\frac{\partial \lx_2(\phi) }{\partial \phi}
~~~~~~~~~~~~
A'=\frac{1}{3} \lx_2(\phi).
\label{bound2} \eeq
By imposing these conditions, we have only to solve
eqs. (\ref{eoma1})--(\ref{eoma3}), neglecting the 
$\delta$-function contributions.

The three equations (\ref{eoma1})--(\ref{eoma3}) are not independent, as
they are related through the Bianchi identities. We look for a solution of
eqs. (\ref{eoma1}), (\ref{eoma3}), which automatically satisfies 
eq. (\ref{eoma2}). Our ansatz for the metric, eq. (\ref{metr1}), indicates
that we should expect a fine-tuning for the existence of 
a static solution \cite{dewolfe}. The reason is that our choice of
four-dimensional Minkowski metric $\eta_{ij}$ requires the vanishing of
the effective cosmological constant on the branes. 

A simple way to understand the fine-tuning is the following:
We can substitute $A'$ as given by eq. (\ref{eoma3}) into 
eq. (\ref{eoma1}). 
Without loss of generality we choose the negative root for $A'$ and
$A(y=0)=0$. We
assume that our low-energy 
Universe corresponds to the positive-tension brane located
at $y=0$ (with $\lx_1(\phi)>0$.). This means that $A<0$ in the bulk.
Eq. (\ref{eoma1}) now becomes a
non-linear second-order differential equation for $\phi(y)$, whose
solution requires two boundary conditions. These are obtained
by substituting eqs. (\ref{bound1}) into eq. (\ref{eoma3}). 
The resulting algebraic equation in general has a discrete number
of solutions that give the allowed values
of $\phi$ at the location of the first brane. 
For each of them the corresponding 
value of $\phi'$ is given by the first of eqs. (\ref{bound1}).
Let us denote generically these solutions by $(\phi_1$, $\phi_1')$.
Now we can integrate eq. (\ref{eoma1}), with $A'$ expressed
in terms of eq. (\ref{eoma3}) and the initial conditions
$\phi(0)=\phi_1$, $\phi'(0)=\phi_1'$. The resulting trajectory 
$(\phi(y),\phi'(y))$ determines the form of the field and the
metric in the bulk. 

In analogy with above, the subsitution of the conditions 
(\ref{bound2}) into eq. (\ref{eoma3}) leads to a discrete 
number of possible values of $\phi$ and $\phi'$ at the location
of the second brane. Let us denote them generically by $(\phi_2,\phi'_2)$.
The fine-tuning is now apparent: The trajectory 
$(\phi(y),\phi'(y))$ must pass through $(\phi_2,\phi'_2)$.
This can be achieved only through a careful choice of 
$\lx_2(\phi)$ (assuming that $\lx_1(\phi)$, $V(\phi)$ are chosen
arbitrarily). However, a possible change of $\lx_1(\phi)$, through
a phase transition on the first brane for example, destabilizes the
solution. The trajectory corresponding to the new initial condition
$(\tilde{\phi}_1,\tilde{\phi}'_1$) does not pass through
$(\phi_2,\phi'_2)$. Some unknown mechanism must
modify the tension $\lx_2(\phi)$ of the second brane 
for a new static solution to exist.

However, there is another possibility:
A new static configuration can still 
exist without modification of $\lx_2(\phi)$
if the 
new boundary conditions $(\tilde{\phi}_1,\tilde{\phi}'_1$) lie on the
initial trajectory $(\phi(y),\phi'(y))$. Then the new solution of eqs.
(\ref{eoma1}), (\ref{eoma3}) is the part of the original one between
$(\tilde{\phi}_1,\tilde{\phi}'_1$) and $(\phi_2,\phi'_2)$.
Physically it corresponds to a small displacement of the positive-tension
brane in a way that the second brane remains unaffected. 
Since the values of $(\tilde{\phi}_1,\tilde{\phi}'_1$) are determined
by the initial function $\lx_1(\phi)$ and its change through a phase
transition, it seems that this scenario is not possible in general.
In the following we discuss how it may work.

\paragraph{A first attempt:}
We start by assuming an initial configuration with a metric given by
eq. (\ref{metr1}). As we explained above this requires an initial fine-tuning
of the brane tensions $\lx_1(\phi)$, $\lx_2(\phi)$. We do not attempt to
address this issue in this work, 
even though we comment on its possible resolution later on.
We are concerned with the requirement of a new fine-tuning every time
$\lx_1(\phi)$ changes. We consider a small change $\lx_1(\phi) \to
\lx_1(\phi) + c(\phi)$, with $|c(\phi)| \ll 1$. There are no constraints
on the form of the function $c(\phi)$, apart from the assumption that it
is small. If we restore the dimensions of the brane tension 
(see eq. (\ref{action0})), our assumption is that $|c(\phi)| \ll (2 M)^4$ 
for the relavant values of $\phi$.

For the new brane tension $\lx_1(\phi) + c(\phi)$, 
the boundary conditions (\ref{bound1}) 
when substituted into eq. (\ref{eoma3})  
lead to a new algebraic equation for $\phi$. 
We denote the solution of this equation by
$\phit_1=\phi_1+\done$, where $\phit_1$ is
the original solution when the tension is given by 
$\lx_1(\phi)$. Assuming $\done = {\cal O} (c(\phi_1))$ and
keeping terms up to order $c(\phi_1)$, we find
\be
\frac{\partial}{\partial \phi}
\left[ V -\frac{1}{8} \left( \frac{\partial \lx_1}{\partial \phi}\right)^2
+ \frac{1}{3} \lx_1^2 \right](\phi_1)\,\, \done=
-\frac{2}{3} \lx_1(\phi_1)c(\phi_1)
+\frac{1}{4} \frac{\partial \lx_1}{\partial \phi}(\phi_1)
\frac{\partial c}{\partial \phi}(\phi_1)+ {\cal O}(c^2(\phi_1)).
\label{dphi1} \ee
For the field derivative we find from the first of eqs. (\ref{bound1}) 
\be
\done'= \frac{1}{2} \frac{\partial^2 \lx_1}{\partial \phi^2}(\phi_1)\,\, \done
+ \frac{1}{2} \frac{\partial c}{\partial \phi} (\phi_1)
+ {\cal O}(c^2(\phi_1)).
\label{dphi1p} \ee
The second of the boundary conditions (\ref{bound1}) provides the new value of
$A'$, which we do not need explicitly\footnote{For 
this section $A$ may be eliminated
as an independent variable. $A'$ can be seen as  
a function of $\phi$ and $\phi'$ defined in
eq. (\ref{eoma3}).}.

As we discussed above, we are looking for modifications of the brane tension
that lead to a new solution lying on the initial trajectory from
$(\phi_1,\phi'_1)$ to $(\phi_2,\phi'_2)$.
We consider this trajectory at a small distance $\dy= {\cal O} (c(\phi_1))$
from the initial location of the positive-tension
brane. By integrating the differential equation (\ref{eoma1})
up to order $\dy$, we find
\beq
\delta \phi'(y)=\phi''(0) \dy + {\cal O}(y^2)
=& \left[ -4 A'(\phi_1) \phi'_1 + 
\frac{\partial V}{\partial \phi}(\phi_1) \right] \dy + {\cal O}(c^2(\phi_1))
\nonumber \\
=&
\left[ \frac{1}{3} \frac{\partial\lx_1^2}{\partial \phi}(\phi_1)+ 
\frac{\partial V}{\partial \phi}(\phi_1) \right] \dy
+ {\cal O}(c^2(\phi_1))
\label{dphi2} \\
\delta \phi (y)=\phi'(0) \dy + {\cal O}(y^2)=& \frac{1}{2} 
\frac{\partial \lx_1}{\partial \phi}(\phi_1) \dy+ {\cal O}(c^2(\phi_1)).
\label{dphi2p}
\eeq

 \begin{figure}[t]
 \centerline{\epsfig{figure=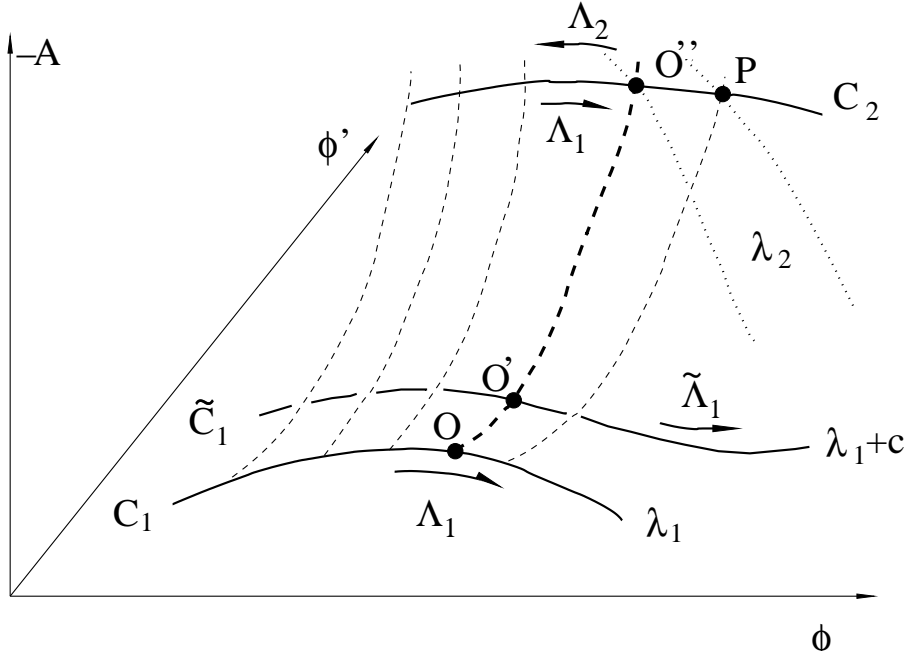,width=12cm}}
 \caption{
 The y-dependence of $\phi$, $\phi'$ and $A'$ near the 
 negative-tension brane.}
 \label{fig2}
 \end{figure}

We would like to identify $\delta \phi(y)$ and $\delta \phi'(y)$
with $\done$ and $\done'$ respectively. The parameter $\dy$ can be adjusted
so as to achieve part of our purpose. 
By solving eq. (\ref{dphi2p}) for $\dy$ and substituting into 
eq. (\ref{dphi2}), we can express $\delta\phi'(y)$ in terms of 
$\delta\phi(y)$.
If we require that the resulting expression be consistent with  
eq. (\ref{dphi1p}) (after the identification 
of  $\delta \phi(y)$, $\delta \phi'(y)$ with $\done$, $\done'$)
we find
\be
\frac{\partial}{\partial \phi}
\left[ V -\frac{1}{8} \left( \frac{\partial \lx_1}{\partial \phi}\right)^2
+ \frac{1}{3} \lx_1^2 \right](\phi_1)\,\, \done =
\frac{1}{4} \frac{\partial \lx_1}{\partial \phi}(\phi_1)
\frac{\partial c}{\partial \phi}(\phi_1)+ {\cal O}(c^2(\phi_1)).
\label{sol1} \ee
For the last expression to be combatible with 
(\ref{dphi1}) we expect a constraint on 
$\lx_1(\phi_1), c(\phi_1)$.
This turns out to be simply
$\lx_1(\phi_1)\, c(\phi_1)=0$ up to order $c(\phi_1)$.
As we would like to keep the form of $c(\phi)$ arbitrary, we are led to 
the condition $\lx_1(\phi_1)=0$. The interpretation is the following:
If the brane is initially located at a position where the value of the
bulk field is such that the brane tension vanishes, subsequent small
modifications of the brane tension can be absorbed in small displacements of
the brane with the metric retaining its four-dimensional Minkowski form.
We emphasize that it is not necessary for $\lx_1(\phi)$ to vanish for
any $\phi$. On the contrary, it may be of order 1 in general. Only the
presence of a zero is necessary. This is possible because the constraint
we derived is independent of the derivatives of $\lx_1(\phi)$, contrary to
the naive expectation.

There are two unsatisfactory elements in our solution: Firstly,
it seems inconsistent to introduce the brane tension as a
$\delta$-function source in
Einstein's equations and then require it to vanish. Notice, however, that
the derivatives of $\lx_1$ at $\phi_1$ are not constrained to be zero. Instead
they may be of order 1.
Secondly, it is questionable if a static solution can be obtained when 
we take into account corrections of order $c^2(\phi_1)$. In order to
resolve these issues we need to expand the framework of our discussion
to include the possibility of more general metrics.
More specifically, we consider
the possibility that the four-dimensional ($y=$const.) part of the metric 
may have (anti-)deSitter form.

\paragraph{Non-zero cosmological constant:}
We consider the ansatz \cite{dewolfe}
\be
ds^2 = e^{2 A(y)}\, g_{ij} \,dx^i dx^j - dy^2,
\label{metr2} \ee
with
\beq
dS_4:~~~~
g_{ij}dx^i dx^j=&dt^2-e^{2\sqrt{\Lx}t}\left( dx_1^2+dx_2^2+dx_3^2 \right)
\label{ds4} \\
AdS_4:~~~~
g_{ij}dx^i dx^j=&e^{-2\sqrt{-\Lx}x_3}\left( dt^2-dx_1^2-dx_2^2 \right)-dx_3^2.
\label{ads4} \eeq
Positive (negative) values of $\Lx$ correspond to deSitter (anti-deSitter)
four-dimensional metrics. 
There is an arbitrariness in the choice of $\Lx$ and $A$, since we can always
set $A=0$ at some point $y$ through a redefinition of the coordinate
$t$ or $x_3$. Only the
combination $A(y)-\ln|\Lx|/2$ is physically relevant.
We remove this ambiguity by 
setting $A(y=0)=0$.
The presence of a non-zero effective cosmological constant $\Lx$
leads to the replacement of eq. (\ref{eoma3}) by \cite{dewolfe}
\be
A'^2 -\Lx e^{-2A}= -\frac{1}{3} V(\phi) + \frac{1}{6} \phi'^2.
\label{eomb3}
\ee
The boundary conditions (\ref{bound1}), (\ref{bound2}) remain unaffected.
Eqs. (\ref{bound1}), (\ref{eomb3}) then demonstrate 
that the value of $\Lx$ is
determined by the mis-match at the location of the first brane
between the brane tension and the
potential of the bulk field that plays the role of the bulk cosmological
constant. The limit $\Lx \to 0$ reproduces the Minkowski metric we considered
earlier.

Within this framework, no fine-tuning is required. General choices of 
$\lx_1(\phi)$, $\lx_2(\phi)$ are expected to lead to a solution with some value
of $\Lx$ \footnote
{It is still possible that no solution exists
for certain choices of $\lx_1(\phi)$, $\lx_2(\phi)$. Our intention is
to show that there are large continuously connected
families of $\lx_1(\phi)$, $\lx_2(\phi)$ for which solutions can be found.
}. 
There is a graphic way to see this: Substitution of 
eqs. (\ref{bound1}) into eq. (\ref{eomb3}) with $\Lx=\Lx_1$ and 
$A_1=0$ gives the value
$\phi_1$ at the location of the positive-tension brane as a function 
of $\Lx_1$.
The first of eqs. (\ref{bound1}) gives $\phi'_1$. The set of initial
conditions forms a curve $C_1$ on the $A=0$ plane, parametrized by $\Lx_1$. 
We assume
that $\Lx_1$ grows in the direction of the arrow in fig. 1.
For given $\Lx_1$, 
the initial conditions $(\phi_1,\phi_1',0)$ result in a 
unique solution for eqs. (\ref{eoma1}), (\ref{eomb3}). This corresponds to
a trajectory in $(\phi,\phi',A)$ space. Varying $\Lx_1$ results in a surface
formed by the various trajectories.

At the location of the second brane, eq. (\ref{eomb3}) must be satisfied
after substitution of eqs. (\ref{bound2}). For given $\Lx=\Lx_2$, the 
possible values $(\phi_2,\phi'_2,A_2)$ form a curve in $(\phi,\phi',A)$ space.
In general, this curve meets the surface of trajectories at some point 
$P$. The 
trajectory going through $P$ corresponds to a value $\Lx_1$ that is
not necessarily equal to $\Lx_2$. However, identification of 
$\Lx_1$ and $\Lx_2$ is expected  
to be possible in general through variation of $\Lx_2$. The 
point $P$ moves with changing $\Lx_2$, in a way that it creates a curve $C_2$
on
the trajectory surface. On $C_2$ trajectories characterized by $\Lx_1$
meet boundary conditions characterized by $\Lx_2$.
For large families of functions $\lx_2(\phi)$,
the identification of $\Lx_1$ and $\Lx_2$ should be possible at
some point on $C_2$  without the necessity of fine-tuning.  
For example, one may consider some function $\lx_2(\phi)$ for which
$\Lx_2$ increases on $C_2$ in the direction indicated by the arrow 
in fig. 1, opposite to the direction of increase of $\Lx_1$. 
The location of $P$ with $\Lx_1=\Lx_2=\Lx$ determines the trajectory 
that corresponds to a static solution of eqs. (\ref{eoma1}), (\ref{eomb3})
under the boundary conditions (\ref{bound1}), (\ref{bound2}).

The presence of the term proportional to $\Lx$ in eq. (\ref{eomb3})
modifies the the solutions relative to the case with
four-dimensional Minkowski metric. 
The nature of the change 
can be seen by solving eq. (\ref{eomb3}) for
a flat potential
$V(\phi)=-\bar{\Lx}$=const. and $\phi$=const.
For $0< |\Lx| \ll 1$ one finds
that $A'$ remains constant (as in the case with $\Lx=0$), 
but then quickly diverges or goes to zero
for $\Lx >0$ or $\Lx <0$ respectively, at 
a distance 
\be
R_1 \simeq -\frac{\sqrt{3}}{2\sqrt{\bar{\Lx}}} \ln |\Lx|
= {\cal O} \left(\left| \ln  |\Lx|  \right| \right)
\label{div} \ee
from the positive-tension brane \cite{dewolfe}.
The negative-tension brane must exist at $y=R < R_1$ in both cases.
For a $y$-dependent $\phi$, one expects that
the solutions with $\Lx=0$  will not be modified significantly
if $|A|$
is sufficiently small for the second term to be negligible relative
to the first one in the lhs of eq. (\ref{eomb3}). This implies that
the second brane must be located at a distance 
$y= R < R_1 \sim {\cal O} \left(\left| \ln  |\Lx|  \right| \right)$.
This is confirmed by numerical studies.
If the negative-tension 
brane in the solution with $\Lx=0$ was 
located at $R > R_1 $, the solution for $\Lx \not= 0$ must
change drastically in order to accomodate the second brane
much closer to the first one than before \cite{far}.
This observation may provide a link between the cosmological constant 
and the brane location. Configurations with branes far apart seem to be
possible only 
if the effective cosmological constant is exponentially small. 

\paragraph{Small cosmological constant:}
We now return to the problem of finding a static configuration after
the tension of the first brane has changed from $\lx_1(\phi)$ to
$\lx_1(\phi)+c(\phi)$. 
With the new brane tension the boundary conditions
at $y=0$ are given by a new curve $\Ct_1$ on the $A=0$ plane. 
Let us consider some point $(\phit_1(\Lt_1),\phit'_1(\Lt_1))$
on $\Ct_1$  (denoted by $O'$) 
close to
the point $(\phi_1(\Lx_1=0),\phi'_1(\Lx_1=0))$ 
on $C_1$ (denoted by $O$).
Writing $\phit_1=\phi_1+\done$ and repeating the calculation that led
to eqs. (\ref{dphi1}), (\ref{dphi1p}) we find
\be
\frac{\partial}{\partial \phi}
\left[ V -\frac{1}{8} \left( \frac{\partial \lx_1}{\partial \phi}\right)^2
+ \frac{1}{3} \lx_1^2 \right](\phi_1)\,\, \done - 3 \Lt_1=
-\frac{2}{3} \lx_1(\phi_1)c(\phi_1)
+\frac{1}{4} \frac{\partial \lx_1}{\partial \phi}(\phi_1)
\frac{\partial c}{\partial \phi}(\phi_1)+ {\cal O}(c^2(\phi_1))
\label{dphi1t} \ee
and
\be
\done'=~ \frac{1}{2} \frac{\partial^2 \lx_1}{\partial \phi^2}(\phi_1)\,\, \done
+ \frac{1}{2} \frac{\partial c}{\partial \phi} (\phi_1)
+ {\cal O}(c^2(\phi_1)).
\label{dphi1pt} \ee

Let us consider the variations of $(\phi,\phi')$ on the
trajectory that starts at 
the point $\Lx_1=0$ on $C_1$.
They are given by eqs. (\ref{dphi2}), (\ref{dphi2p}). 
It is possible now to identify them with $(\done,\done')$ of eqs.
(\ref{dphi1t}), (\ref{dphi1pt}) if
\be
\Lt_1= \frac{2}{9} \lx_1(\phi_1)c(\phi_1) + {\cal O}(c^2(\phi_1)).
\label{cosm} \ee
For $\lx_1(\phi_1) = {\cal O} (c(\phi_1))$,
we obtain
$\Lt_1={\cal O}(c^2(\phi_1))$. 
This means that
the trajectory that starts at
the point $\Lx_1=0$ on $C_1$ passes through\footnote{
More precisely, the trajectory does not go exactly through $\Ct_1$.
As $A'(y=0)=-\lx_1(\phi_1)/3$
and $\dy={\cal O}(c(\phi_1))$, we
have $A(y=\dy) ={\cal O}(c^2(\phi_1))$.
However, this small deviation of $A$ from 0 can be neglected
in our considerations. This is obvious from eq.~(\ref{eomb3}), if
we assume $\Lx={\cal O}(c^2(\phi_1))$. A shift of $A$ at some point by 
${\cal O}(c^2(\phi_1))$ gives an effect ${\cal O}(c^4(\phi_1))$.}
the curve $\Ct_1$ at some point 
with $\Lt_1={\cal O}(c^2(\phi_1))$.
We also know where this trajectory ends:
on the curve $C_2$ at $\Lx_2=0$. (This is the initial fine-tuning
that we assumed to have been achieved.)

If we concentrate only on the part of the trajectory from
$\Ct_1$ to $C_2$, we have the situation we analyzed earlier. A
slight mis-match between $\Lt_1$ and $\Lx_2$. But, as we argued before,
there should be a nearby trajectory for which $\Lt_1$ and $\Lx_2$ can be
matched, especially if $\Lt_1$ and $\Lx_2$ grow in opposite directions
on $C_2$. 
For the case of interest $\Lt_1={\cal O}(c^2(\phi_1))$
and $\Lx_2=0$, this trajectory is expected to have 
$\Lx_f={\cal O}(c^2(\phi_1))$.

The upshot of this complicated reasoning is that we identified
a static solution of Einstein's equations with tensions 
$\lx_1(\phi)+c(\phi)$ for the first brane and
$\lx_2(\phi)$ for the second. It is a solution with an
effective four-dimensional constant $\Lx$.
Contrary to expectations, $\Lx$ is not of order $c(\phi_1)$, but of
order $c^2(\phi_1)$. The reason is that the location of the first
brane has been shifted by an amount $\dy={\cal O}(c(\phi_1))$, so
as to absorb the leading effect from the change of the brane tension.
The necessary condition is   
$\lx_1(\phi_1) = {\cal O} (c(\phi_1))$ or smaller.
In order not to change the positivity of the tension for any sign of
$c(\phi_1)$,  it is 
preferable to take
$\lx_1(\phi_1)$ somewhat larger than $|c(\phi_1)|$.

\paragraph{Discussion:}

The basic objective of our approach was to compensate a possible
change in a brane tension with a slight modification of the 
two-brane configuration in a way that keeps the effective cosmological
constant $\Lx$ small. We considered
variations of the positive tension 
$\lx_1(\phi)$ of the first 
brane. The reason is that we identify our Universe with the positive-tension
brane, while we view the negative-tension one as a regulator that cuts
off possible singularities in the solutions of 
Einstein's equations.  We allowed small arbitrary
changes $c(\phi)$ of $\lx_1(\phi)$, while we kept fixed the form of the
negative tension $\lx_2(\phi)$ and the potential $V(\phi)$ of the bulk field. 
We view $\lx_1(\phi)$, $\lx_2(\phi)$ and $V(\phi)$ as effective 
low-energy quantities. In particular, $\lx_1(\phi)$ represents the vacuum
energy on the first brane, induced by the fields localized on it (such as the
Standard Model fields) and their interactions with
the bulk field. The changes $c(\phi)$ originate in
variations of the characteristic
energy scale on the first brane, or possible phase transitions
on the brane. We assumed that, apart from isolated points where it 
may approach zero,
$\lx_1(\phi)$ is of order
1 in units of the fundamental Planck's constant, while $|c(\phi)| \ll 1$.

We started from an initial configuration with zero effective cosmological 
constant. This requires an initial fine-tuning of 
$\lx_1(\phi)$ and  $\lx_2(\phi)$ for which we do not have a convincing
explanation. We can only speculate that, since the two branes
can exist far apart only if $\Lx$ is exponentially
small, the fine-tuning 
may be a consequence of the initial location of the branes.
Another possibility is that the initial configuration has exact supersymmetry
that forces the effective cosmological constant to be zero.
This situation should correspond to a large energy scale on the first brane. 
When the energy scale is lowered, supersymmetry gets broken and
an effective cosmological constant may appear.
Our concern was the destabilization of the initial configuration every time
$\lx_1(\phi)$ changes. We looked for possible new static configurations, 
similar to the initial one. We did not address the question of 
the evolution of the system from one configuration to the other.
This requires a time-dependent solution of Einstein's equations, with
ansatzes for the metric much more general than the ones we employed.
The technical difficulties involved in obtaining such solutions
are a significant obstacle in this direction.

We demonstrated in a graphic way the known fact that, for 
general $\lx_1(\phi)$, $\lx_2(\phi)$, a static configuration exists
with some value of the cosmological constant. Consequently, every
change $c(\phi)$ of $\lx_1(\phi)$ from its initial fine-tuned form 
results in a new static configuration with non-zero $\Lx$.
In general, one expects $\Lx = {\cal O}(c(\phi_1))$, where $\phi_1$ is
the value of the bulk field at the initial location of the first brane.
We showed that this is not always the case.
Our main result is that one can have $\Lx = {\cal O}(c^2(\phi_1))$.
The only requirement is that $\phi_1$ is near a zero of
$\lx_1(\phi)$, so that $\lx_1(\phi_1) = {\cal O}(c(\phi_1))$.
We point out that the derivatives of $\lx_1(\phi)$ at $\phi_1$ are
in general of order 1. If $\lx_1(\phi)$ can become negative for a certain
range of $\phi$ we assume
that $\lx_1(\phi_1)$ is somewhat larger than $|c(\phi_1)|$,   
so that the positivity of the tension is maintained for any sign of
$c(\phi_1)$. Another possibility is that $\lx_1(\phi)$
has minimum near $\phi_1$, where $\lx_1(\phi)$ and $c(\phi)$ are
comparable.

The implications of our result for a possible 
realistic scenario are interesting. In order to be
more specific, we return to dimensionful quantities. 
We take the fundamental constant $M$ defined in
the beginning to be $M={\cal O} (10^{19}$ GeV). For the four-dimensional
Planck's constant we expect \cite{dewolfe} 
\be \frac{M^2_4}{M^2} \sim \int_0^{R} dr\, e^{2A(r)} = {\cal O} (1).
\label{fourd} \ee
Recent observations are consistent with a cosmological constant of the
order of the critical density of the Universe \cite{obsc}
\be
\frac{\Lx}{M^4} = {\cal O} \left(10^{-120} \right).  
\label{obscosm} \ee
If we assume an initial two-brane configuration with zero effective
cosmological constant, subsequent changes in the tension of the 
first brane are consistent with the above constraints if
\be
c(\phi_1)={\cal O} \left( 10^{-60} M^4 \right)
={\cal O} \left( \left( 10{\rm~ TeV}\right)^4 \right).
\label{cbound} \ee

Let us consider the possibility that for a
certain value $\phi_1$ of the bulk field
there is an ultraviolet cutoff of order 10 TeV
for the vacuum energy associated with the fields of the first brane.
If at the initial location of the brane $\phi \simeq \phi_1$,
then $\lx(\phi_1)$ is of order $\left( 10{\rm~ TeV}\right)^4$.
Our results imply that if the effective
cosmological constant was zero at the beginning,
any subsequent 
modification $c(\phi)$ of the brane tension may result in a
cosmological constant consistent with the observational data
as long as 
$c(\phi) \sim 
{\cal O} \left( \left( 10{\rm~ TeV}\right)^4 \right)$.
In this scenario, smaller modifications of the tension, such as those
caused by 
cosmological phase transitions at scales below 10 TeV, do not 
modify the cosmological constant substantially. 
All phase transitions predicted by
known physics (such as the electroweak or the QCD phase transitions)
fall in this category. The nature of the cutoff cannot be specified by
our considerations. It is possible that supersymmetry provides
the necessary mechanism for the cancellation of quantum contributions
to the energy density at energy scales above 10 TeV.

Before concluding, we would like to emphasize that the mechanism
we presented should be viewed only as a simple example of how 
a non-trivial topology can ameliorate the cosmological constant problem.
We find that its most important merit relative to alternative proposals
within the same framework \cite{adks} is the absence of singularities and 
strong assumptions about the form of the interactions of the 
brane fields with the bulk field. Our initial ansatz for the effective action
of the system is general and our only assumption about the changes
of the brane tension is that they are small in units of Planck's constant. 
Moreover, our scenario allows for a non-zero cosmological constant of
the right order of magnitude \cite{obsc}.

As a final remark, we point out that in 
eqs. (\ref{dphi1}), (\ref{sol1}), (\ref{dphi1t}) we assumed that the bulk
potential cannot be given identically by
\be
V(\phi)= 
\frac{1}{8} \left[ \frac{\partial \lx_1(\phi)}{\partial \phi}\right]^2
- \frac{1}{3} \left[ \lx_1 (\phi) \right]^2. 
\label{final} \ee
This interesting expression could be favoured by gauged supergravity.
(The example of ref. \cite{ellw}, based on the construction of ref. 
\cite{lukas}, fulfills this relation.)
If eq. (\ref{final}) holds, any value of $\phi$ can be taken as 
the value of the bulk field at the location of the first brane. As a result,
a continuous range of possible values for the tension of the first brane
could lead to solutions with the same tension for the
negative brane.
This can be viewed as an improvement with respect to the case
of arbitrary $V(\phi)$, $\lx_1(\phi)$, for which only a discrete number
of $\phi$ values is possible. However, our approach is
much more general. We allow for arbitrary changes of the form of 
$\lx_1(\phi)$, not just its value. 
Such changes would invalidate the relation (\ref{final}).

\vspace{0.5cm}
\noindent
{\bf Acknowledgements}: 
I would like to thank E. Kiritsis, A. Krause, A. Strumia and T. Tomaras
for useful discussions.
This research was supported in part by the European Commission under 
the RTN programs HPRN--CT--2000--00122, HPRN--CT--2000--00131 and
HPRN--CT--2000--00148.

\end{document}